\pdfoutput=1
\documentclass[11pt]{article}
\usepackage[T1]{fontenc}
\usepackage[utf8]{inputenc}
\usepackage{lmodern}
\usepackage{microtype}
\usepackage{amsmath,amssymb,amsthm}
\usepackage{mathtools}
\IfFileExists{dsfont.sty}{\usepackage{dsfont}}{}
\providecommand{\mathds}[1]{\mathbf{#1}}
\usepackage[left=3.18cm,right=3.18cm,top=2.54cm,bottom=2.54cm]{geometry}
\usepackage{setspace}
\usepackage{natbib}
\setcitestyle{round}
\usepackage{graphicx}
\usepackage{booktabs}
\usepackage{float}
\usepackage{xcolor}
\usepackage[colorlinks,linkcolor=blue,anchorcolor=blue,citecolor=blue]{hyperref}

\newtheorem{theorem}{Theorem}
\newtheorem{proposition}{Proposition}
\theoremstyle{definition}
\newtheorem{example}{Example}

\begin{document}

\title{Systemic Correlation by Simple Counting}

\author{Peter Reinhard Hansen\textsuperscript{a,}\thanks{Corresponding author. Email address: hansen@unc.edu.}
\ and Zhengtai Zheng\textsuperscript{b}\\[1.5ex]
{\small \textsuperscript{a}University of North Carolina at Chapel Hill, Chapel Hill, NC 27599, USA}\\
{\small \textsuperscript{b}University of Chicago, Chicago, IL 60637, USA}}

\date{November 2026}

\maketitle

\begin{abstract}
We show that the systemic correlation among the constituents of a stock index can be measured by simple counting of sign concordances, which bypasses the need to estimate volatilities. The daily measure depends only on the numbers of positive and negative stock returns, and is closely related to the sign concordance between the stocks and the index. We model these quantities with a simple score-driven model, and apply it to daily returns for the constituents of the S\&P 500 index, 1990–2025. The sign-based measure is shown to closely track the average pairwise Pearson correlation.
\end{abstract}



\section{Introduction}

The average pairwise correlation among stocks, or systemic correlation,
is widely used to summarize the dependence in equity markets. It predicts aggregate stock returns \citep{PolletWilson:2010},
and correlation risk appears to be priced \citep{KrishnanPetkovaRitchken:2009,DriessenMaenhoutVilkov:2009}.
Conventional estimators of the systemic correlation, such as the dynamic
equicorrelation model of \citet{EngleKelly:2012}, require a volatility
model for every stock, which is demanding when the number of stocks
is large and introduces a source of estimation and misspecification
error.

Comovement can also be measured by counting. The synchronicity measure
of \citet{MorckYeungYu:2000} is the fraction of stocks that move in
the majority direction. The sign concordance of two returns measures how often they have the
same sign, and we show that the average sign concordance across all
pairs of stocks on a given day is an increasing function of this fraction.
Thus, a simple count delivers an average of pairwise dependence measures,
and the mapping between sign concordance and correlation expresses
it on the conventional correlation scale. The measure is robust to
outliers and requires no volatility estimates, which is convenient
because volatility varies substantially across days and within the
trading day, see, e.g., \citet{HansenLuo-RobustCorr:2023}.

Next, we relate the average pairwise sign concordance, the systemic
concordance, to the index concordance, defined as the average sign
concordance between each stock and the equal-weighted index. The daily
systemic statistic is an exact function of the squared daily index
statistic, and this yields a lower bound for the systemic concordance
that requires no distributional assumptions (Proposition \ref{prop:Signs}).
We establish an analogous result for Pearson correlations in Theorem
\ref{thm:SystemicIndex}. 

We model the daily time series of both concordance measures with the
binomial model of \citet{HansenSchmidtblaicher:Hesitancy}, which belongs
to the class of score-driven models, see \citet{CrealKoopmanLucas:2013}
and \citet{Harvey:2013}. In an empirical analysis of the constituents
of the S\&P 500 index over the period 1990--2025, we find that the
sign-based measure closely tracks the average pairwise Pearson correlation.

\section{Systemic and index sign concordance\label{sec:Concordance}}

For two returns $X$ and $Y$, let $\tau=\mathbb{E}[\operatorname{sgn}(XY)]$
denote their sign concordance, which coincides with the quadrant correlation
\citep{Sheppard:1898,Blomqvist:1950} when both medians are zero. If $(X,Y)$ is elliptically distributed
around zero, with correlation $\rho$ and $\mathbb{P}((X,Y)=(0,0))=0$,
then
\begin{equation}
\rho=\rho(\tau)\equiv\sin\left(\tfrac{\pi}{2}\tau\right),\label{eq:greiner}
\end{equation}
see \citet{HansenLuo-RobustCorr:2023} for details. The sign concordance
is invariant to the scale of each return. Thus, when each return is
multiplied by its own positive and possibly time-varying volatility,
$\rho(\tau)$ continues to give the correlation of $(X,Y)$. In contrast,
the Pearson correlation of the scaled returns is closer to zero when
the volatilities are predictable and not proportional. Outside this class the mapping
need not recover the Pearson
correlation, and even under ellipticity, applying (\ref{eq:greiner})
to an average of concordances need not give the average correlation,
because the transformation is nonlinear. We therefore take sign
concordances as the quantities of interest and use $\rho(\tau)$ to express them
on the conventional correlation scale.

Consider the returns on $n$ stocks on a given day, $r_{1},\ldots,r_{n}$,
and let $R=n^{-1}\sum_{i}r_{i}$ be the equal-weighted index. Let $U$
and $D$ denote the numbers of positive and negative returns, and let
$B=(U-D)/M$ denote the breadth, where $M=U+D$. The systemic statistic
is the average of $\operatorname{sgn}(r_{i}r_{j})$ over the pairs with a
nonzero product,\footnote{Zero
returns are rare after decimalization in 2001 (0.8\% of stock-days),
but were common in the first part of our sample, when prices moved
in eighths of a dollar (12\% of stock-days in 1990--1996). Counting pairs with a zero return as $\tau=0$ lowers the
average filtered systemic correlation from 0.23 to 0.19 in 1990--1996
and by less than 0.01 after decimalization.} which is given by
\begin{equation}
\hat{\tau}_{\bullet}=1-\frac{4UD}{M(M-1)}=\frac{MB^{2}-1}{M-1}\in\left[-\frac{1}{M-1},1\right].\label{eq:count}
\end{equation}
The bounds coincide with those of an equicorrelation for even $M$,
see \citet{ArchakovHansen:CanonicalBlockMatrix}. The fraction of stocks
moving in the majority direction is $F=\max(U,D)/M$, such that $|B|=2F-1$,
and $\hat{\tau}_{\bullet}$ is an increasing function of the synchronicity
measure of \citet{MorckYeungYu:2000}.

The index statistic is the corresponding average for the index, $\hat{\tau}_{\star}=M^{-1}\sum_{i}\operatorname{sgn}(r_{i}R)=\operatorname{sgn}(R)B$,
which is the breadth signed by the index return. The systemic statistic
is invariant to rescaling the individual returns, whereas the index
statistic is not, because rescaling can change the sign of $R$.

Weights are simple to accommodate. With weights $w_{i}\geq0$ that
sum to one over the stocks with nonzero returns, the breadth is replaced
by the weighted breadth, $B_{\mathrm{w}}=\sum_{i}w_{i}\operatorname{sgn}(r_{i})$,
and (\ref{eq:count}) becomes $\hat{\tau}_{\bullet}=(B_{\mathrm{w}}^{2}-\sum_{i}w_{i}^{2})/(1-\sum_{i}w_{i}^{2})$.

We define the \emph{systemic concordance} and the \emph{index concordance}
as the conditional means of the two statistics given the number of
nonzero returns, $\tau_{\bullet}=\mathbb{E}[\hat{\tau}_{\bullet}|M]$
and $\tau_{\star}=\mathbb{E}[\hat{\tau}_{\star}|M]$. Without zero
returns, $M=n$, and $\tau_{\bullet}$ is the average of the pairwise
concordances, $\mathbb{E}[\operatorname{sgn}(r_{i}r_{j})]$, whereas $\tau_{\star}$
is the average of the concordances between each stock and the index,
$\mathbb{E}[\operatorname{sgn}(r_{i}R)]$.
\begin{proposition}
\label{prop:Signs}Suppose that $M\geq2$ and $R\neq0$. Then $\hat{\tau}_{\bullet}=(M\hat{\tau}_{\star}^{2}-1)/(M-1)$,
and
\[
\frac{1+(M-1)\tau_{\bullet}}{M}-\tau_{\star}^{2}=\operatorname{var}(\hat{\tau}_{\star}|M),\qquad\text{so that}\qquad\tau_{\bullet}\geq\frac{M\tau_{\star}^{2}-1}{M-1}.
\]
\end{proposition}
The first part follows from (\ref{eq:count}), because $\hat{\tau}_{\star}^{2}=B^{2}$,
and the second by taking conditional expectations. The result requires
neither ellipticity nor equal volatilities, and it accommodates zero
returns. The identity has the following simple interpretation: The
index concordance, $\tau_{\star}$, is the conditional mean of the
index statistic, and $[1+(M-1)\tau_{\bullet}]/M$ is its conditional
second moment, such that the left-hand side is its conditional variance.
Thus, the systemic concordance exceeds the lower bound by more when
the index statistic is more variable. Note that $\hat{\tau}_{\star}$
and $B$ can differ in sign, because a majority of stocks can rise
on a day when the equal-weighted index falls.

\subsection{The analogous result for Pearson correlations}

Under ellipticity, each pairwise sign concordance maps to the
corresponding Pearson correlation through (\ref{eq:greiner}). Next,
we derive a result for Pearson correlations that is analogous to Proposition
\ref{prop:Signs}. Let $\sigma_{i}$ and $\rho_{ij}$ denote the standard deviations and
correlations of the returns, and let $w_{i}\geq0$ be weights with
$\sum_{i}w_{i}=1$. The systemic correlation is $\rho_{\bullet}=\sum_{i<j}w_{ij}\rho_{ij}$,
with $w_{ij}=w_{i}w_{j}/\sum_{k<l}w_{k}w_{l}$, and the index correlation
is $\varrho_{\star}=\sum_{i}w_{i}\operatorname{corr}(r_{i},R)$, where
the index is now the weighted average, $R=\sum_{i}w_{i}r_{i}$.
\begin{theorem}
\label{thm:SystemicIndex}Let $w=(w_{1},\ldots,w_{n})^{\prime}$,
$\Lambda_{\sigma}=\operatorname{diag}(\sigma_{1},\ldots,\sigma_{n})$,
and $C$ denote the correlation matrix of $(r_{1},\ldots,r_{n})^{\prime}$,
with $0<\sigma_{i}<\infty$, $w^{\prime}w<1$, and $\operatorname{var}(R)>0$.
Then, with $\rho_{\mathrm{w}}=w^{\prime}Cw$ and $\tilde{R}=\sum_{i}w_{i}r_{i}/\sigma_{i}$,
\[
\rho_{\bullet}=\frac{\rho_{\mathrm{w}}-w^{\prime}w}{1-w^{\prime}w},\qquad\varrho_{\star}=\frac{w^{\prime}C\Lambda_{\sigma}w}{\sqrt{w^{\prime}\Lambda_{\sigma}C\Lambda_{\sigma}w}},\qquad\rho_{\mathrm{w}}-\varrho_{\star}^{2}=\operatorname{var}\left(\tilde{R}-\tfrac{\operatorname{cov}(\tilde{R},R)}{\operatorname{var}(R)}R\right)\geq0,
\]
so that $\rho_{\bullet}\geq(\varrho_{\star}^{2}-w^{\prime}w)/(1-w^{\prime}w)$.

(i) If $\sigma_{1}=\cdots=\sigma_{n}$, then $\rho_{\mathrm{w}}=\varrho_{\star}^{2}$.

(ii) If $\rho_{ij}=r$ for all $i\neq j$, then $\rho_{\mathrm{w}}-\varrho_{\star}^{2}\leq(1-r)\sum_{i}w_{i}^{2}(1-\sigma_{i}/\bar{\sigma}_{\mathrm{w}})^{2}$,
where $\bar{\sigma}_{\mathrm{w}}=\sum_{i}w_{i}\sigma_{i}$.
\end{theorem}
The proof is in the Appendix. The first identity has the same form
as the weighted version of (\ref{eq:count}), with $\rho_{\mathrm{w}}$
in place of $B_{\mathrm{w}}^{2}$. The gap, $\rho_{\mathrm{w}}-\varrho_{\star}^{2}$,
is the residual variance from regressing the weighted sum of the standardized
returns, $\tilde{R}$, on the index. It vanishes when all volatilities are
equal and, by (ii), it is of order $\sum_{i}w_{i}^{2}$ when all correlations
are equal and the volatilities are bounded away from zero and infinity. However, the bound in Theorem \ref{thm:SystemicIndex} does not carry
over to the sign-based measures after they are mapped to the correlation
scale with (\ref{eq:greiner}). In Example \ref{S:ex:Gaussian} in the Supplement, $\rho(\tau_{\bullet})=0.276$,
which is below the bound $[\rho(\tau_{\star})^{2}-1/n]/(1-1/n)=0.340$.

\section{A score-driven model for the daily counts}

The daily correlation measures are noisy, but their underlying
population measures are persistent.
This motivates filtering, which estimates the current population measure
from a weighted average of past daily measures. We add a time index,
$t$, to the quantities in Section \ref{sec:Concordance}, and let $\{\mathcal{F}_{t}\}$
denote the natural filtration, such that $\mathcal{F}_{t}$ is generated
by the returns up to and including day $t$. Let $\mathcal{G}_{t}=\mathcal{F}_{t-1}\vee\sigma(M_{t})$
be the $\sigma$-algebra generated by $\mathcal{F}_{t-1}$ and $M_{t}$,
such that $\mathcal{F}_{t-1}\subset\mathcal{G}_{t}\subset\mathcal{F}_{t}$.
The systemic and index concordances on day $t$ are $\tau_{\bullet,t}=\mathbb{E}[\hat{\tau}_{\bullet,t}|\mathcal{G}_{t}]$
and $\tau_{\star,t}=\mathbb{E}[\hat{\tau}_{\star,t}|\mathcal{G}_{t}]$,
and Proposition \ref{prop:Signs} holds conditionally on $\mathcal{G}_{t}$
by the same argument. Let $X_{t}$ denote the number of concordant
pairs among the $N_{t}=M_{t}(M_{t}-1)/2$ pairs with a nonzero product
on day $t$, such that $\hat{q}_{t}=X_{t}/N_{t}=(1+\hat{\tau}_{\bullet,t})/2$.

We adopt the score-driven binomial model for the conditional mean,
$\mathbb{E}[X_{t}|\mathcal{G}_{t}]=N_{t}q_{t}$, where $q_{t}=(1+\tau_{\bullet,t})/2$.
The model is parametrized with the log-odds, $\eta_{t}=\log[q_{t}/(1-q_{t})]$,
which evolve according to
\begin{equation}
\eta_{t+1}=\omega+\beta\eta_{t}+\alpha s_{t},\qquad s_{t}=\frac{\hat{q}_{t}-q_{t}}{\sqrt{q_{t}(1-q_{t})}},\label{eq:ScoreModel}
\end{equation}
where $s_{t}$ is a normalized version of the score of the binomial
log-likelihood, $\ell_{t}=X_{t}\log q_{t}+(N_{t}-X_{t})\log(1-q_{t})$,
whose derivative with respect to $\eta_{t}$ is $N_{t}(\hat{q}_{t}-q_{t})$.
The score-driven update is intuitive, because it raises $q_{t}$ after
a day with more concordant pairs than expected and lowers it after
a day with fewer, which is the same principle that underlies exponential
smoothing. The index concordance is modeled in the same way, with
$X_{\star,t}=\sum_{i}\mathds{1}_{\{r_{i,t}R_{t}>0\}}$ and $N_{\star,t}=M_{t}$.
By Proposition \ref{prop:Signs}, the two filters track the first and
second conditional moments of the index statistic, $\hat{\tau}_{\star,t}$.

We estimate $\theta=(\omega,\alpha,\beta)$ by maximizing $\sum_{t}\ell_{t}(\theta)$,
with $\eta_{1}$ fixed at the log-odds of the concordance rate in the
first two months of the sample. The binomial likelihood is a quasi-likelihood,
because the pairwise indicators share stocks and respond to common
news, which makes $X_{t}$ overdispersed relative to a binomial count.
Under standard identification and stability conditions, the estimator
remains consistent if the conditional mean is correctly specified,
see \citet{GourierouxMonfortTrognon:1984}. However, conventional (non-robust)
standard errors overstate the precision. We therefore report robust
standard errors, with the long-run variance of the scores estimated
with the Parzen kernel and the bandwidth chosen by the plug-in rule
of \citet{Andrews:1991}.

\section{Empirical illustration\label{sec:Empirical}}

We use daily returns, including dividends, for the constituents of
the S\&P 500 index from CRSP. The sample period is January 1990 to
December 2025, which has 9,067 trading days. Index membership is determined
day by day from the CRSP index constituent file, and we drop the missing
returns, which are 143 of about 4.5 million stock-days. We do not
center the returns, because daily medians are close to zero. Centering
each return at its median within the calendar year lowers the annual
sign-based measure reported below by 0.006 on average, and changes
it by at most 0.02.

The estimates in Table \ref{tab:Estimates} show that both measures
are highly persistent. The half-life of a shock to the log-odds is
about 55 trading days for the systemic measure and 75 days for the
index measure, and it is about 35 days for both in the sample after
2000. The overdispersion is large, as the variance of the daily concordance
rate around its filtered value is about 6,600 times the binomial variance
for the systemic count and about 40 times for the index count. This
means that a day provides the same information as about 18 and 13
independent Bernoulli trials would, respectively, instead of $N_{t}\approx120{,}000$
and $M_{t}\approx500$ trials, which is the main reason that the robust
standard errors are about 120 and 8 times larger than the conventional
ones. The Supplement estimates the model with exact likelihoods for
$X_{t}$, derived from models with a common correlation. These models
imply an overdispersion of the magnitude found here, and their filtered
series have a correlation of 0.997 with the series reported below.
\begin{table}[H]
\caption{Estimates of the score-driven model\label{tab:Estimates}}
\begin{center}
\begin{tabular*}{\textwidth}{@{\extracolsep{\fill}}lcccc}
\toprule
 & \multicolumn{2}{c}{1990--2025} & \multicolumn{2}{c}{2001--2025}\tabularnewline
\cmidrule{2-3}\cmidrule{4-5}
 & Systemic & Index & Systemic & Index\tabularnewline
\midrule
$\hat{\omega}$ & 0.006 & 0.008 & 0.010 & 0.017\tabularnewline
 & (0.001) & (0.002) & (0.002) & (0.004)\tabularnewline
$\hat{\alpha}$ & 0.114 & 0.090 & 0.126 & 0.107\tabularnewline
 & (0.014) & (0.012) & (0.015) & (0.014)\tabularnewline
$\hat{\beta}$ & 0.988 & 0.991 & 0.980 & 0.981\tabularnewline
 & (0.003) & (0.003) & (0.005) & (0.005)\tabularnewline
Half-life (days) & 55 & 75 & 34 & 37\tabularnewline
\bottomrule
\end{tabular*}
\end{center}
{\small Note: Quasi-maximum likelihood estimates of (\ref{eq:ScoreModel})
for the systemic count, $X_{t}$, and the index count, $X_{\star,t}$,
with $\eta_{1}$ fixed at the value implied by the first two months
of each sample. Robust standard errors in parentheses, with a Parzen-kernel
long-run variance and the bandwidth of \citet{Andrews:1991}, between
6 and 7 days. The half-life is $\log(0.5)/\log\hat{\beta}$, computed
from the unrounded estimates.}
\end{table}

\begin{figure}[H]
\begin{center}
\includegraphics[width=1\textwidth]{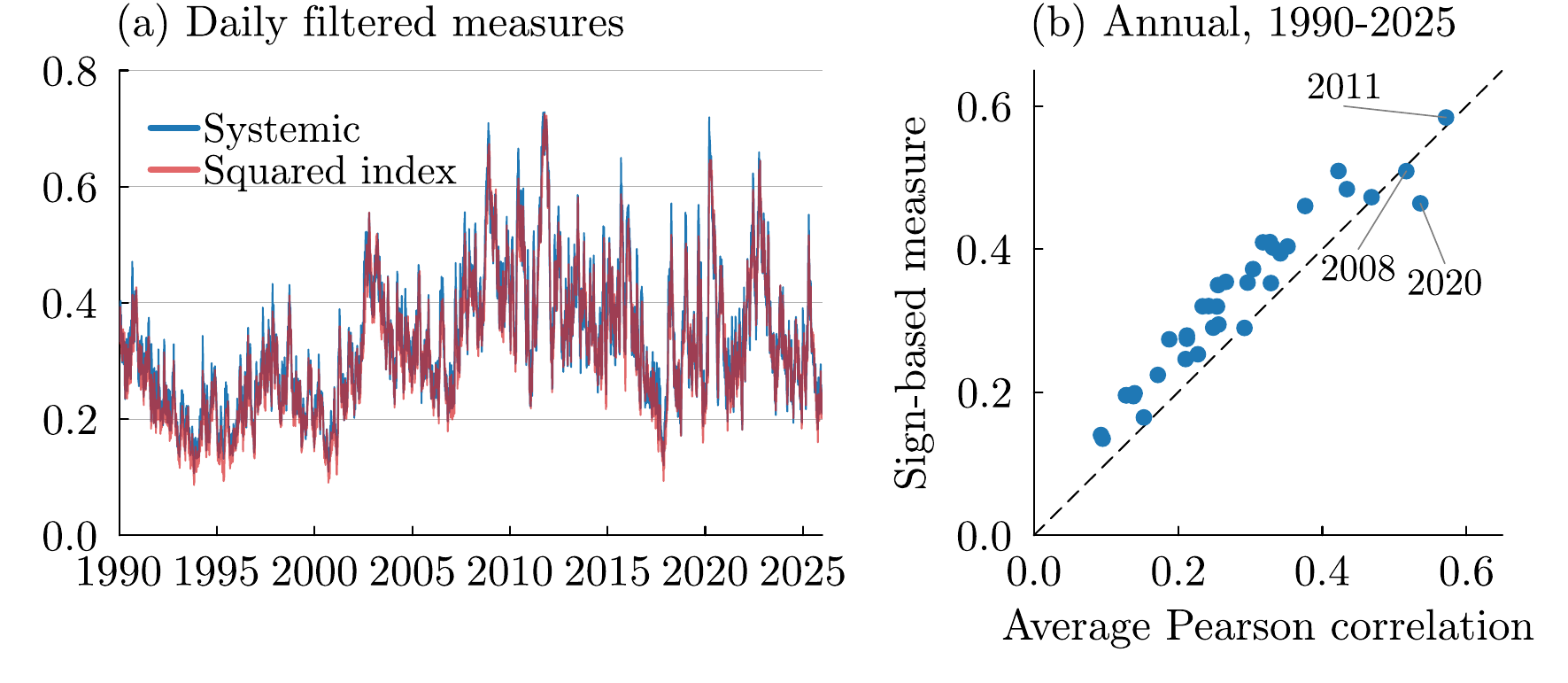}
\caption{S\&P 500 constituents, 1990--2025. (a) The filtered systemic and
squared index measures on the correlation scale, $\rho(\tau_{\bullet,t})$
and $\rho(\tau_{\star,t})^{2}$, from the score-driven model (\ref{eq:ScoreModel}).
(b) Annual sign-based measure against the average pairwise Pearson
correlation, with the 45-degree line; the three years with the highest
average Pearson correlation are labeled.\label{fig:SystemicIndex}}
\end{center}
\end{figure}

\begin{figure}[H]
\begin{center}
\includegraphics[width=1\textwidth]{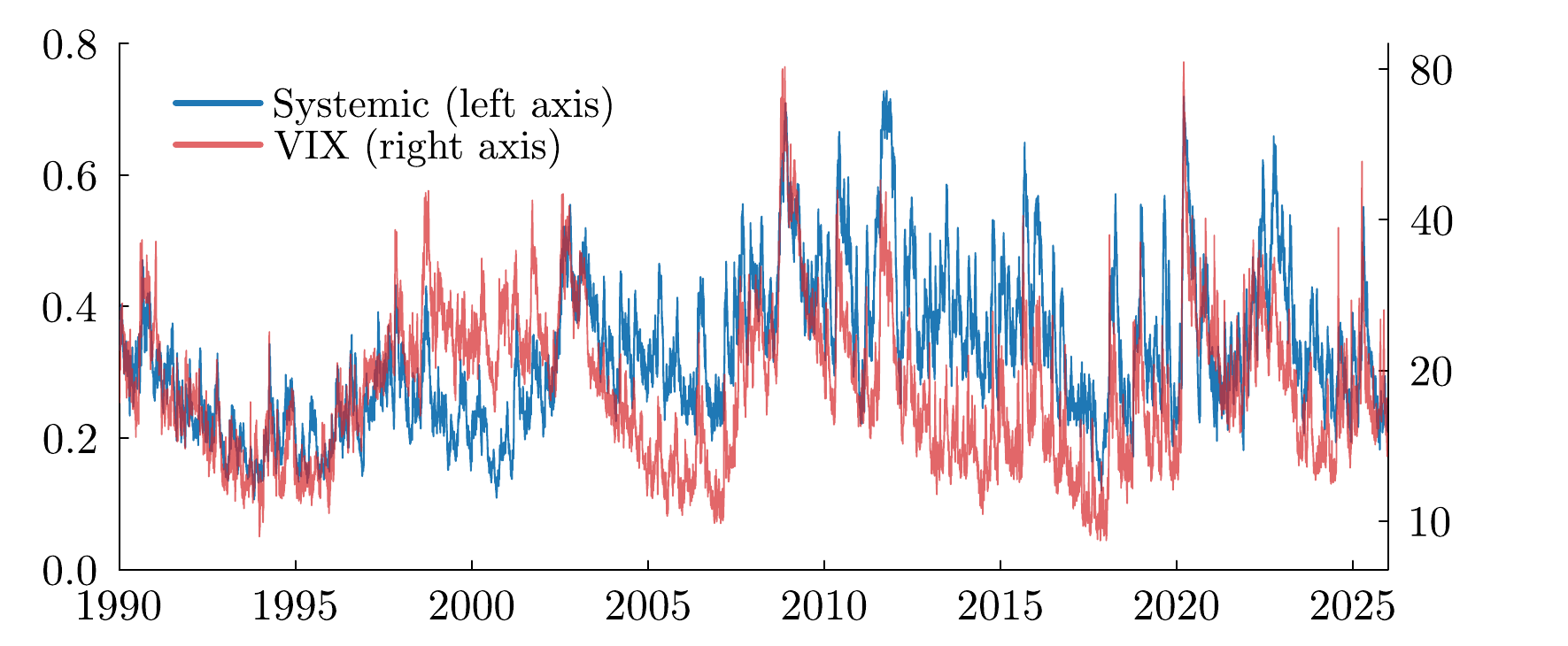}
\caption{The filtered systemic correlation, $\rho(\tau_{\bullet,t})$
(left axis), and the CBOE Volatility Index, VIX (right axis, log
scale), daily,
January 1990 to December 2025. The correlation between the two series
is 0.54 in levels and 0.64 in three-month changes. VIX values before September 2003 are back-calculated by Cboe.\label{fig:SystemicVIX}}
\end{center}
\end{figure}

Figure \ref{fig:SystemicIndex}(a) plots the two filtered measures
on the correlation scale, $\rho(\tau_{\bullet,t})$ and $\rho(\tau_{\star,t})^{2}$.
They are very close, with a correlation of 0.99, and the systemic
measure is higher by 0.010 on average (0.98 and 0.009 after 2000).
The two series agree partly by construction, because both come from
$\hat{\tau}_{\star,t}$. A useful benchmark is centered jointly Gaussian
returns with equal variances and common correlation $r$, for which
$\rho(\tau_{\bullet})=r$ and $\rho(\tau_{\star})^{2}=r+(1-r)/n$,
such that the squared index measure is slightly higher in the benchmark,
whereas the systemic measure is higher in the data, which is explained
in the Supplement. Both measures
rise sharply in periods of market stress, and the systemic measure
peaks in November 2008
(the global financial crisis), October 2011 (the European sovereign
debt crisis), and March 2020 (the onset of the COVID-19 pandemic). Figure \ref{fig:SystemicVIX} plots
the systemic measure with the VIX, and their correlation is 0.54 at
the dates plotted.
The two series are related mainly through their slow movements, because
the correlation between their three-month changes is 0.64, whereas
it is 0.16 for their daily changes, where the change in the filtered
measure that is caused by the returns on day $t$ is matched with the
change in the VIX on day $t$.

To assess how well the measure matches conventional correlations,
Figure \ref{fig:SystemicIndex}(b) compares annual estimates. For
each calendar year, we use the stocks that were in the index, with
a return, on every trading day of the year, and compute three quantities.
The first is the sign-based measure, which applies the mapping (\ref{eq:greiner})
to the equal-weighted average of the pairwise estimates, $\hat{\tau}_{ij}$,
each computed over the days on which both returns are nonzero. The
second applies (\ref{eq:greiner}) to each $\hat{\tau}_{ij}$ before
averaging, and the third is the average pairwise Pearson correlation,
computed from all days. The first exceeds the second by 0.005 on average
and never by more than 0.01, such that it matters little whether (\ref{eq:greiner})
is applied before or after averaging. Across the 36 years, the sign-based
measure has a correlation of 0.96 with the average Pearson correlation
in levels and 0.94 in annual changes, which shows that the agreement
cannot be ascribed to spurious correlation between two persistent
series. However, the sign-based measure is higher in 33 of the years,
by 0.05 on average (0.33 against 0.28; 0.38 against 0.33 after 2000),
the exceptions being 2008, 2020, and 2025. A difference in this direction
is expected when the correlations are constant within the year, because
time-varying volatility then attenuates the Pearson correlation toward
zero, unless the volatilities of the two stocks are proportional,
see \citet{HansenLuo-RobustCorr:2023}. If the correlations are higher
on days with high volatility, then the Pearson correlation gives these
days more weight, which works in the opposite direction. Averaging the daily statistic
(\ref{eq:count}) over each year, instead of the pairwise estimates,
gives nearly the same result, with a correlation of 0.96 with the
average Pearson correlation and an average difference of 0.05.

Figure \ref{fig:SystemicIndex}(b) labels the three crisis years,
2008, 2011, and 2020, which have the highest average Pearson correlations.
In 2008 and 2011 the two measures differ by about 0.01, whereas in
2020 the sign-based measure is 0.07 below the Pearson correlation.
In each of these years, a few days with large returns of the same
sign account for much of the Pearson correlation. Table \ref{tab:Extreme}
shows that excluding the ten days with the largest absolute index
returns lowers the average Pearson correlation by 0.09 to 0.14, but
the sign-based measure by only about 0.03. In 2020, excluding five
days reduces the difference between the two measures from 0.07 to
less than 0.002. This sensitivity is expected,
because the Pearson correlation has an unbounded influence function,
whereas that of the sign concordance is bounded, see, e.g., \citet{HansenLuo-RobustCorr:2023}.
Across all 36 years, excluding
these ten days lowers the sign-based measure by between 0.02 and 0.04,
but the Pearson correlation by up to 0.14. Thus, the sign-based measure
captures the variation in the average Pearson correlation well, at
a somewhat higher level, and it depends far less on a handful of days.
\begin{table}[H]
\caption{Annual measures with and without the days with the largest
absolute index returns\label{tab:Extreme}}
\begin{center}
\small
\begin{tabular*}{\textwidth}{@{\extracolsep{\fill}}lcccccc}
\toprule
 & \multicolumn{3}{c}{Sign-based measure} & \multicolumn{3}{c}{Average Pearson correlation}\tabularnewline
\cmidrule{2-4}\cmidrule{5-7}
 & All days & Excl.\ 5 & Excl.\ 10 & All days & Excl.\ 5 & Excl.\ 10\tabularnewline
\midrule
2008 & 0.509 & 0.494 & 0.477 & 0.517 & 0.457 & 0.412\tabularnewline
2011 & 0.584 & 0.569 & 0.555 & 0.572 & 0.514 & 0.480\tabularnewline
2020 & 0.464 & 0.446 & 0.430 & 0.536 & 0.448 & 0.394\tabularnewline
\bottomrule
\end{tabular*}
\end{center}
{\small Note: Annual estimates as in Figure \ref{fig:SystemicIndex}(b),
computed from all trading days of the year and after excluding the
5 or 10 days with the largest absolute return on the equal-weighted
index. The larger changes in the Pearson correlation reflect its unbounded
influence function.}
\end{table}

\section{Conclusion}

The numbers of positive and negative stock returns each day yield
the average
pairwise sign concordance, a measure of systemic dependence that requires no
volatility model and is an exact function of the daily synchronicity
count. It is tied to the index concordance by a distribution-free
identity, and a score-driven binomial filter with robust inference
tracks both. On the correlation scale the measure follows the average
Pearson correlation closely, at a level about 0.05 higher, so it is
best interpreted as a sign concordance expressed on the correlation scale. The method can also be
applied to intraday returns, as in \citet{HansenLuo-RobustCorr:2023}.
Natural extensions include market-capitalization weights, comparisons
with DECO-type models, and block structures, such as sectors, where
the average concordance between the stocks in two different blocks,
$g$ and $h$, is the product of their breadths, $B_{g}B_{h}$, see \citet{ArchakovHansen:CanonicalBlockMatrix}
for block correlation matrices.

\section*{Declaration of generative AI and AI-assisted technologies in the manuscript preparation process}

During the preparation of this work, the authors used ChatGPT (OpenAI)
and Claude (Anthropic) in order to obtain feedback on drafts, to assist
with drafting and editing the text, to write and check the computer
code for the empirical analysis, and to assist with deriving and verifying
the results in the Supplement. After using these tools, the authors
reviewed and edited the content as needed and take full responsibility
for the content of the paper.

\section*{Acknowledgements}

This paper is, in part, based on Zhengtai Zheng's Senior Honors Thesis
in Economics at UNC, ``Time-varying Correlation and Portfolio Optimization
in the U.S. Stock Market.'' We thank Klara Peter for helpful comments.
This research did not receive any specific grant from funding agencies
in the public, commercial, or not-for-profit sectors.

\bibliographystyle{apalike}
\bibliography{prh}

\appendix

\section*{Appendix: Proof of Theorem \ref{thm:SystemicIndex}}

Let $w_{\bullet}=\sum_{1\leq i<j\leq n}w_{i}w_{j}$. Then
\[
\rho_{\bullet}=\frac{1}{w_{\bullet}}\sum_{i<j}w_{i}w_{j}\rho_{ij}=\frac{1}{2w_{\bullet}}\left(\sum_{i,j}w_{i}w_{j}\rho_{ij}-\sum_{i}w_{i}^{2}\right)=\frac{1}{1-w^{\prime}w}\left(\rho_{\mathrm{w}}-w^{\prime}w\right),
\]
where $\rho_{\mathrm{w}}=\sum_{i,j}w_{i}w_{j}\rho_{ij}=w^{\prime}Cw$. Next,
\[
\operatorname{var}(R)=\sum_{i=1}^{n}\sum_{j=1}^{n}w_{i}w_{j}\operatorname{cov}(r_{i},r_{j})=w^{\prime}\Lambda_{\sigma}C\Lambda_{\sigma}w,
\]
and $\operatorname{var}(r_{i})=\sigma_{i}^{2}$, such that
\[
\varrho_{\star}=\sum_{i=1}^{n}\frac{w_{i}\operatorname{cov}(r_{i},R)}{\sqrt{\operatorname{var}(R)}\sigma_{i}}=\frac{w^{\prime}C\Lambda_{\sigma}w}{\sqrt{w^{\prime}\Lambda_{\sigma}C\Lambda_{\sigma}w}}.
\]
Define $\tilde{R}=\sum_{i=1}^{n}\frac{w_{i}r_{i}}{\sigma_{i}}$. Then,
\[
|w^{\prime}C\Lambda_{\sigma}w|=|\operatorname{cov}(\tilde{R},R)|\leq\sqrt{\operatorname{var}(\tilde{R})\operatorname{var}(R)}=\sqrt{(w^{\prime}Cw)(w^{\prime}\Lambda_{\sigma}C\Lambda_{\sigma}w)},
\]
by Cauchy--Schwarz. Squaring both sides and rearranging proves the
inequality $\varrho_{\star}^{2}\leq\rho_{\mathrm{w}}$. More precisely, $\operatorname{var}(\tilde{R})=\rho_{\mathrm{w}}$
and $\operatorname{cov}(\tilde{R},R)^{2}/\operatorname{var}(R)=\varrho_{\star}^{2}$, so $\rho_{\mathrm{w}}-\varrho_{\star}^{2}$
is the variance of the residual from projecting $\tilde{R}$ on $R$.

In case (i), with equal volatilities, we have $\Lambda_{\sigma}=\sigma I_{n}$
and $\operatorname{corr}(\tilde{R},R)=1$, which implies $\rho_{\mathrm{w}}=\varrho_{\star}^{2}$.

For (ii), the projection residual minimizes $\operatorname{var}(\tilde{R}-aR)$
over $a$, so $\rho_{\mathrm{w}}-\varrho_{\star}^{2}\leq\operatorname{var}(\tilde{R}-R/\bar{\sigma}_{\mathrm{w}})$.
Now $\tilde{R}-R/\bar{\sigma}_{\mathrm{w}}=\sum_{i}c_{i}r_{i}/\sigma_{i}$
with $c_{i}=w_{i}(1-\sigma_{i}/\bar{\sigma}_{\mathrm{w}})$ and $\sum_{i}c_{i}=0$,
so with $\rho_{ij}=r$ its variance is $(1-r)\sum_{i}c_{i}^{2}+r(\sum_{i}c_{i})^{2}=(1-r)\sum_{i}c_{i}^{2}$.
\qed

\clearpage

\setcounter{section}{0}
\setcounter{equation}{0}
\setcounter{table}{0}
\setcounter{figure}{0}
\renewcommand{\thesection}{S.\arabic{section}}
\renewcommand{\theequation}{S.\arabic{equation}}
\renewcommand{\thetable}{S.\arabic{table}}
\renewcommand{\thefigure}{S.\arabic{figure}}
\renewcommand{\theHsection}{S.\arabic{section}}
\renewcommand{\theHequation}{S.\arabic{equation}}
\renewcommand{\theHtable}{S.\arabic{table}}
\renewcommand{\theHfigure}{S.\arabic{figure}}

\section*{Supplement}

This supplement has three parts. Section \ref{S:sec:Example} presents
the example that is referred to in Section \ref{sec:Concordance}
of the paper. Section \ref{S:sec:Exact} derives the exact distribution
of the systemic count, $X_{t}$, for models in which the signs of
the returns are exchangeable, and estimates two such models. Section
\ref{S:sec:Difference} explains why the systemic measure is higher
than the squared index measure in Figure \ref{fig:SystemicIndex}(a).
Theorems,
equations, tables, and figures in the paper are referred to by their
numbers, whereas those in this supplement have the prefix ``S''.

\section{An example for Theorem \ref{thm:SystemicIndex}\label{S:sec:Example}}

The following example shows that the inequality in Theorem \ref{thm:SystemicIndex}
need not hold for the sign concordances after they are mapped to the
correlation scale.

\begin{example}\label{S:ex:Gaussian}
Let $n=500$ returns be centered and jointly Gaussian with unit variances
and $w_{i}=1/n$. The first 250 have pairwise correlation 0.9, and
the remaining 250 are independent of each other and of the first group.
Then $\rho_{\bullet}=0.2245$ and $\varrho_{\star}^{2}=0.2261$, which
satisfy Theorem \ref{thm:SystemicIndex}(i), because $(\varrho_{\star}^{2}-1/n)/(1-1/n)=0.2245$.
The concordances are $\tau_{\bullet}=0.178$ and $\tau_{\star}=0.397$,
which satisfy Proposition \ref{prop:Signs} with $M=n$, because $(n\tau_{\star}^{2}-1)/(n-1)=0.156$.
On the correlation scale, however, $\rho(\tau_{\bullet})=0.276$ and
$\rho(\tau_{\star})^{2}=0.341$, such that $\rho(\tau_{\bullet})$
is below $[\rho(\tau_{\star})^{2}-1/n]/(1-1/n)=0.340$.
\end{example}

\section{Exact likelihoods for the systemic count\label{S:sec:Exact}}

The paper estimates the score-driven model (\ref{eq:ScoreModel})
with a binomial quasi-likelihood. This likelihood has the correct
conditional mean, but it greatly understates the variance of $X_{t}$.
In this section, we derive the exact distribution of $X_{t}$ for
models with exchangeable signs, and we estimate (\ref{eq:ScoreModel})
with two likelihoods of this type. The purpose is to examine whether the
empirical results in the paper depend on the binomial quasi-likelihood,
and to explain the overdispersion.

\subsection{The distribution of the count\label{S:sec:Distribution}}

Conditional on $\mathcal{G}_{t}$, there are $M_{t}$ nonzero returns,
of which $U_{t}$ are positive and $D_{t}=M_{t}-U_{t}$ are negative,
and the systemic count is $X_{t}=N_{t}-U_{t}D_{t}$. The function
$u\mapsto u(M_{t}-u)$ is symmetric about $M_{t}/2$ and strictly
increasing for $u\leq M_{t}/2$. Thus, $X_{t}$ is one-to-one with
the unordered pair $\{U_{t},D_{t}\}$. Let $p_{t}(u)=\mathbb{P}(U_{t}=u|\mathcal{G}_{t})$
for $u=0,\ldots,M_{t}$. It follows that
\begin{equation}
\mathbb{P}\left(X_{t}=N_{t}-u(M_{t}-u)|\mathcal{G}_{t}\right)=\begin{cases}
p_{t}(u)+p_{t}(M_{t}-u) & \text{for }u<M_{t}/2,\\
p_{t}(u) & \text{for }u=M_{t}/2.
\end{cases}\label{S:eq:ExactPmf}
\end{equation}
The exact likelihood for $X_{t}$ is therefore available whenever
the distribution of the number of positive returns is available.

Suppose that the signs of the $M_{t}$ returns are independent conditional
on a latent variable, $P_{t}\in[0,1]$, and that each return is positive
with probability $P_{t}$, such that $U_{t}|P_{t}\sim\operatorname{Bin}(M_{t},P_{t})$.
This is the structure that arises when the dependence between the
returns is caused by a single common factor. Two returns have the
same sign with probability $P_{t}^{2}+(1-P_{t})^{2}$, conditional
on $P_{t}$, and $\mathbb{E}[U_{t}D_{t}|P_{t}]=M_{t}(M_{t}-1)P_{t}(1-P_{t})$.
Thus,
\begin{equation}
\mathbb{E}[X_{t}|\mathcal{G}_{t}]=N_{t}q_{t},\qquad\text{with}\quad q_{t}=1-2\mathbb{E}[P_{t}(1-P_{t})|\mathcal{G}_{t}],\label{S:eq:Mean}
\end{equation}
which is the conditional mean assumed in the paper, and this holds
for any distribution of $P_{t}$. It also follows that $\tau_{\bullet,t}=2q_{t}-1=(1-2\mathbb{E}[P_{t}|\mathcal{G}_{t}])^{2}+4\operatorname{var}(P_{t}|\mathcal{G}_{t})$,
such that the systemic concordance is nonnegative in these models.

The count is not informative about asymmetry in the distribution of
$P_{t}$. If $P_{t}$ is replaced by a variable that equals $P_{t}$
or $1-P_{t}$ with equal probability, then $p_{t}(u)$ is replaced
by $\tfrac{1}{2}[p_{t}(u)+p_{t}(M_{t}-u)]$, and the distribution
of $X_{t}$ in (\ref{S:eq:ExactPmf}) is unchanged. Thus, it is without
loss of generality to take the distribution of $P_{t}$ to be symmetric
about $\tfrac{1}{2}$ in a likelihood for $X_{t}$.

The distribution of $P_{t}$ is important for the variance. From
$\hat{q}_{t}=1-U_{t}D_{t}/N_{t}$ and $\mathbb{E}[\hat{q}_{t}|P_{t}]=1-2P_{t}(1-P_{t})$,
we have
\begin{equation}
\operatorname{var}(\hat{q}_{t}|\mathcal{G}_{t})=4\operatorname{var}\left(P_{t}(1-P_{t})|\mathcal{G}_{t}\right)+O(M_{t}^{-1}).\label{S:eq:Variance}
\end{equation}
The leading term does not depend on the number of stocks, whereas
the binomial variance, $q_{t}(1-q_{t})/N_{t}$, is of order $M_{t}^{-2}$.
Thus, if $\operatorname{var}(P_{t}(1-P_{t})|\mathcal{G}_{t})>0$,
the overdispersion relative to the binomial distribution is of order
$N_{t}$. This condition is not implied by a positive concordance.
For instance, if $P_{t}$ equals $\tfrac{1}{4}$ or $\tfrac{3}{4}$
with equal probability, then $\tau_{\bullet,t}=\tfrac{1}{4}$ whereas
$P_{t}(1-P_{t})$ is constant, and the overdispersion is $(2M_{t}+1)/5$.
The condition holds for the two models we consider below. When it
holds, we define the \emph{variance-equivalent number of independent
trials} by
\begin{equation}
n_{t}^{\ast}=\frac{q_{t}(1-q_{t})}{4\operatorname{var}\left(P_{t}(1-P_{t})|\mathcal{G}_{t}\right)},\label{S:eq:EffectiveTrials}
\end{equation}
such that the leading term in (\ref{S:eq:Variance}) is the variance
of the average of $n_{t}^{\ast}$ independent Bernoulli trials with
success probability $q_{t}$. This number does not depend on the number
of stocks in the index. It depends on the distribution of $P_{t}(1-P_{t})$,
and it is therefore not determined by $\tau_{\bullet,t}$ alone.

\subsection{Two models}

We consider two distributions for $P_{t}$. Both are symmetric about
$\tfrac{1}{2}$, which is without loss of generality for $X_{t}$,
and both are fully determined by $q_{t}$, such that no parameters
are added to the model.

The first is the \emph{probit-normal} model. Suppose that $r_{i,t}=\sigma_{i,t}(\sqrt{\rho_{t}}f_{t}+\sqrt{1-\rho_{t}}\varepsilon_{i,t})$,
where $f_{t}$ and $\varepsilon_{i,t}$, $i=1,\ldots,M_{t}$, are
independent and standard normal, and the volatilities, $\sigma_{i,t}>0$,
are arbitrary. All pairwise correlations are equal to $\rho_{t}$,
and (\ref{eq:greiner}) holds for every pair, such that $\rho_{t}=\rho(\tau_{\bullet,t})$.
A return is positive when $\varepsilon_{i,t}>-\sqrt{\rho_{t}/(1-\rho_{t})}f_{t}$.
Thus, $P_{t}=\Phi(Y_{t})$ with $Y_{t}\sim N(0,\lambda_{t})$ and
$\lambda_{t}=\rho_{t}/(1-\rho_{t})$, where $\Phi$ denotes the standard
normal distribution function, and
\begin{equation}
p_{t}(u)=\binom{M_{t}}{u}\int_{-\infty}^{\infty}\Phi(y)^{u}[1-\Phi(y)]^{M_{t}-u}\frac{1}{\sqrt{\lambda_{t}}}\phi\left(\frac{y}{\sqrt{\lambda_{t}}}\right)\mathrm{d}y,\label{S:eq:ProbitNormal}
\end{equation}
where $\phi$ is the standard normal density. We evaluate the integral
numerically on a grid with 801 points.

The second is the \emph{beta-binomial} model, where $P_{t}\sim\operatorname{Beta}(a_{t},a_{t})$
and
\begin{equation}
p_{t}(u)=\binom{M_{t}}{u}\frac{\mathrm{B}(u+a_{t},M_{t}-u+a_{t})}{\mathrm{B}(a_{t},a_{t})},\label{S:eq:BetaBinomial}
\end{equation}
with $\mathrm{B}(\cdot,\cdot)$ denoting the beta function. From
(\ref{S:eq:Mean}), we have $q_{t}=(a_{t}+1)/(2a_{t}+1)$, such that
$a_{t}=(1-\tau_{\bullet,t})/(2\tau_{\bullet,t})$. \citet{HarmonEtAl:2015}
fit this distribution, with symmetric parameters, to the annual distribution
of the number of stocks with positive returns. Here we relate the
distribution to the sign concordance and use it as a likelihood for
the score-driven model. The beta-binomial model is convenient, because
the probabilities are available in closed form, and because the variance-equivalent
number of trials in (\ref{S:eq:EffectiveTrials}) simplifies to
\begin{equation}
n_{t}^{\ast}=\frac{(1+\tau_{\bullet,t})(1+2\tau_{\bullet,t})}{2\tau_{\bullet,t}^{2}}.\label{S:eq:EffectiveTrialsBeta}
\end{equation}
For instance, $\tau_{\bullet,t}=0.21$ gives $n_{t}^{\ast}\approx19$.
In both models we have $p_{t}(u)=p_{t}(M_{t}-u)$, such that the
probability in (\ref{S:eq:ExactPmf}) is $2p_{t}(u)$ for $u<M_{t}/2$,
and both models require $q_{t}>\tfrac{1}{2}$.

\subsection{Estimation and results}

We keep the recursion (\ref{eq:ScoreModel}) unchanged, including
the binomial score, $s_{t}$. Thus, $q_{t}$ is the same function
of $\theta$ and past observations as in the paper, and only the likelihood
is different. We estimate $\theta$ by maximizing the exact log-likelihood,
$\sum_{t}\log\mathbb{P}(X_{t}|\mathcal{G}_{t})$, which is given by
(\ref{S:eq:ExactPmf}) in combination with either (\ref{S:eq:ProbitNormal})
or (\ref{S:eq:BetaBinomial}). The sample is January 1990 to December
2025, and $\eta_{1}$ is fixed at the same value as in the paper.
The results are reported in Table \ref{S:tab:Exact}, next to the
quasi-maximum likelihood estimates from Table \ref{tab:Estimates}.

\begin{table}[H]
\caption{Estimates with the binomial quasi-likelihood and two exact likelihoods\label{S:tab:Exact}}
\begin{center}
\small
\begin{tabular*}{\textwidth}{@{\extracolsep{\fill}}lccc}
\toprule
 & Binomial (QML) & Probit-normal & Beta-binomial\tabularnewline
\midrule
$\hat{\omega}$ & 0.0058 & 0.0046 & 0.0045\tabularnewline
 & (0.00001) {[}0.0014{]} & (0.0009) {[}0.0012{]} & (0.0009) {[}0.0011{]}\tabularnewline
$\hat{\alpha}$ & 0.1136 & 0.0865 & 0.0871\tabularnewline
 & (0.0001) {[}0.0142{]} & (0.0086) {[}0.0112{]} & (0.0087) {[}0.0112{]}\tabularnewline
$\hat{\beta}$ & 0.9876 & 0.9886 & 0.9888\tabularnewline
 & (0.00003) {[}0.0030{]} & (0.0023) {[}0.0028{]} & (0.0023) {[}0.0028{]}\tabularnewline
Half-life (days) & 55 & 61 & 62\tabularnewline
\midrule
Log-likelihood, probit-normal & $-48{,}420.7$ & $-48{,}413.5$ & $-48{,}413.5$\tabularnewline
Log-likelihood, beta-binomial & $-48{,}418.8$ & $-48{,}411.9$ & $-48{,}411.9$\tabularnewline
\midrule
Eigenvalues of $A^{-1}B$ & 4,393 & 0.82 & 0.82\tabularnewline
 & 8,913 & 1.25 & 1.23\tabularnewline
 & 14,683 & 1.77 & 1.74\tabularnewline
$\log\det(A^{-1}B)$ & 27.08 & 0.59 & 0.56\tabularnewline
\midrule
Average of $\rho(\tau_{\bullet,t})$ & 0.329 & 0.320 & 0.320\tabularnewline
\bottomrule
\end{tabular*}
\end{center}
{\small Note: Estimates of (\ref{eq:ScoreModel}) for the systemic
count, $X_{t}$, January 1990 to December 2025. Conventional standard
errors, based on the Hessian, are in parentheses, and robust standard
errors are in brackets. The two exact log-likelihoods are evaluated
at the estimates in each column. The matrix $A$ is the negative average
Hessian, and $B$ is the long-run variance of the scores, estimated
with the Parzen kernel and the bandwidth of \citet{Andrews:1991}.}
\end{table}

The estimates with the two exact likelihoods are nearly identical,
and they are similar to the quasi-maximum likelihood estimates. The
estimate of $\beta$ is almost unchanged, whereas the estimate of
$\alpha$ is smaller, 0.087 as opposed to 0.114, which is a difference
of about two robust standard errors. The half-life increases from
55 to about 61 days. The differences have little effect on the filtered
series. On the correlation scale, the filtered systemic measure from
either exact likelihood has a correlation of 0.997 with the series
in Figure \ref{fig:SystemicIndex}(a), the average absolute difference
is 0.011, and the largest difference is below 0.07. The two exact likelihoods
yield filtered series that differ by at most 0.002. The exact log-likelihoods
are about 7 units lower at the quasi-maximum likelihood estimates
than at their maxima, and the beta-binomial log-likelihood is 1.6
units higher than the probit-normal log-likelihood, which is a small
difference between two models with the same number of parameters.
Thus, the empirical results in the paper do not depend on the binomial
quasi-likelihood.

The exact models explain the overdispersion that is reported in the
paper. With the beta-binomial estimates, the average of $\operatorname{var}(\hat{q}_{t}|\mathcal{G}_{t})/[q_{t}(1-q_{t})]$
over the sample is $1/18.8$, where the variance is the one implied
by the model. The corresponding average with $(\hat{q}_{t}-q_{t})^{2}$
in place of the variance is $1/18.6$. Relative to the binomial variance,
the model-implied variance is about 6,400 times larger on average,
and the empirical variance is about 6,500 times larger. Thus, the
beta-binomial model implies about 19 variance-equivalent trials per
day, in agreement with (\ref{S:eq:EffectiveTrialsBeta}), which is
what we observe in the data, and the probit-normal model implies nearly
the same variance. It follows that the large overdispersion is not
evidence against the specification of the conditional mean, because
an overdispersion of order $N_{t}$ is implied by both models, see
(\ref{S:eq:Variance}).

The information matrix equality, $A=B$, holds for a correctly specified
likelihood, see, e.g., \citet{White:1982}, in which case all eigenvalues
of $A^{-1}B$ are one. For the binomial quasi-likelihood the eigenvalues
are between 4,400 and 14,700, which is the reason that the robust
standard errors in Table \ref{tab:Estimates} are 92 to 120 times
larger than the conventional standard errors. For the two exact likelihoods
the eigenvalues are between 0.82 and 1.77, and the robust standard
errors are 20\% to 30\% larger than the conventional ones. The robust
standard errors of the exact maximum likelihood estimators are also
somewhat smaller than those of the quasi-maximum likelihood estimator,
0.011 as opposed to 0.014 for $\hat{\alpha}$.

We do not claim that either of the two exact models is correctly specified.
As a diagnostic, we simulated 200 samples from the estimated beta-binomial
model and computed $\log\det(A^{-1}B)$ for each sample at the parameter
values used in the simulation, with $B$ estimated by the outer product
of the scores. The central 95\% of the simulated values are between
$-0.30$ and $0.33$, whereas the value in the data, which is computed
at the estimated parameters, is 0.57. This is not a formal test, because
the parameters were not estimated in the simulated samples. However,
it suggests that the beta-binomial model does not fully describe the
data, albeit the deviation from the information matrix equality is
small in comparison with that of the binomial quasi-likelihood. We
did not simulate from the probit-normal model, but its eigenvalues
are nearly the same. Two limitations of the exact models are that
both have a common correlation for all pairs of stocks, and that the
recursion uses the binomial score and not the score of the exact likelihood.
For these reasons, robust standard errors are also appropriate for
the exact likelihoods, and the binomial quasi-likelihood with robust
standard errors, which is used in the paper, is a simple alternative
that gives nearly the same filtered series.

The number of positive returns, $U_{t}$, has more information than
the count, because $X_{t}$ does not distinguish between days where
most returns are positive and days where most returns are negative.
Under the probit-normal model, $z_{t}=\Phi^{-1}((U_{t}+\tfrac{1}{2})/(M_{t}+1))/\sqrt{\lambda_{t}}$
is approximately standard normal when $M_{t}$ is large. In the data,
$z_{t}$ has mean 0.03, variance 1.01, skewness $-0.26$, and kurtosis
3.10. The variance and kurtosis do not depend on the sign of $z_{t}$,
and they are close to the values implied by the model. The negative
skewness shows that days with a large share of negative returns are
more pronounced than days with a large share of positive returns.
This asymmetry does not affect the likelihood for $X_{t}$, as explained
in Section \ref{S:sec:Distribution}, but it would be relevant for
a model of $U_{t}$.

\section{The systemic and the squared index measures\label{S:sec:Difference}}

The benchmark in Section \ref{sec:Empirical} of the paper has centered
Gaussian returns with a common correlation, $r$, which is the probit-normal
model in Section \ref{S:sec:Exact}. In the benchmark, the squared
index measure exceeds the systemic measure by $(1-r)/n$, which is
about 0.001. In Figure \ref{fig:SystemicIndex}(a), the systemic measure
is the higher of the two, by 0.010 on average. There are two reasons
for this difference.

First, the index statistic, $\hat{\tau}_{\star,t}=\operatorname{sgn}(R_{t})B_{t}$,
is negative when the return on the index and the majority of the stocks
have opposite signs. This occurs on 5.4\% of the days in our sample.
For comparison, we simulated the benchmark with $M_{t}$ stocks and
the correlation given by the filtered value, $\rho(\tau_{\bullet,t})$,
and the index statistic is negative on 1.6\% of the simulated days.
The average of $\hat{\tau}_{\star,t}$ is 0.376 in the data and 0.387
in the simulated benchmark. About half of this difference is due to
the days with a negative index statistic, because the average of $|B_{t}|-\hat{\tau}_{\star,t}$
is 0.006 in the data and 0.001 in the benchmark.

Second, the correlation on day $t$ need not be known given $\mathcal{G}_{t}$.
Suppose that the returns are as in the benchmark conditional on $r_{t}$,
and that $r_{t}\geq0$ is random given $\mathcal{G}_{t}$. Then $\tau_{\bullet,t}=\mathbb{E}[\tfrac{2}{\pi}\arcsin(r_{t})|\mathcal{G}_{t}]$
and $\tau_{\star,t}=\mathbb{E}[\tfrac{2}{\pi}\arcsin(\sqrt{x_{t}})|\mathcal{G}_{t}]$
with $x_{t}=r_{t}+(1-r_{t})/n$. The function $\arcsin(r)$ is convex
for $r\geq0$, and $\arcsin(\sqrt{x})$ is concave for $x\leq\tfrac{1}{2}$.
It follows by Jensen's inequality that $\rho(\tau_{\bullet,t})\geq\mathbb{E}[r_{t}|\mathcal{G}_{t}]$
and, when $x_{t}\leq\tfrac{1}{2}$, that $\rho(\tau_{\star,t})^{2}\leq\mathbb{E}[x_{t}|\mathcal{G}_{t}]$.
Thus, a random correlation raises the systemic measure and lowers
the squared index measure. However, this effect is small in our data,
because a random correlation also makes the tails of $z_{t}$ heavier.
For instance, if $r_{t}$ has a beta distribution with mean 0.33 and
standard deviation 0.15, then the systemic measure exceeds the squared
index measure by 0.013, which is similar to the difference in Figure
\ref{fig:SystemicIndex}(a), but the kurtosis of $z_{t}$ would be
about 5.1. With a standard deviation of 0.05 the kurtosis is about
3.2, which is close to the value of 3.10 in the data, and the difference
between the two measures is 0.000, as opposed to $-0.001$ when the
correlation is known.

\end{document}